# Artificial intelligence and the internal processes of creativity

Jaan Aru

Institute of Computer Science, University of Tartu

jaan.aru@gmail.com

**Abstract**

Artificial intelligence (AI) systems capable of generating creative outputs are reshaping our understanding of creativity. This shift presents an opportunity for creativity researchers to reevaluate the key components of the creative process. In particular, the advanced capabilities of AI underscore the importance of studying the internal processes of creativity. This paper explores the neurobiological machinery that underlies these internal processes and describes the experiential component of creativity. It is concluded that although the products of artificial and human creativity can be similar, the internal processes are different. The paper also discusses how AI may negatively affect the internal processes of human creativity, such as the development of skills, the integration of knowledge, and the diversity of ideas.

**Natural and artificial creativity**

Recent advances in artificial intelligence push the boundaries of our understanding of creativity. This is perhaps best illustrated by considering the standard definition of creativity (Runco & Jaeger, 2012), which states that "Creativity requires both originality and effectiveness".

The output of present-day generative AI algorithms can clearly satisfy these criteria (Runco, 2023a,b). For instance, one can tell a large language model (such as ChatGPT or Claude) to come up with a haiku that meets these criteria, and it will. Hence, we have to either accept that AI is creative similarly to humans or update this standard definition to distinguish between artificial and human creativity (Runco, 2023a).

Two recent suggestions for updating the standard definition of creativity are relevant in the present context. First, Runco proposed that we should distinguish artificial creativity from natural human creativity by including authenticity and intentionality in the definition (Runco, 2023a). Second, Abraham (2024) proposed that "a creative idea is one that is both novel and satisfying".

Both of these attempts to redefine creativity highlight the importance of internal processes for understanding natural human creativity. For instance, for a novel idea to be satisfying, the system has to be able to feel satisfaction. The experience aspect is essential, as stressed by Morris Stein: "On completing the creative product, the creative individual *experiences a feeling* of satisfaction" (Stein, 1963, italics added). This is one of the key reasons why humans create (Abraham, 2024). Carl Rogers would agree that "we must face the fact that the individual creates primarily because it is satisfying to him" (Rogers, 1954, p. 252, cited in Abraham, 2024).

Hence, to fully understand human creativity and how it differs from artificial creativity, we must more closely look at the inner processes underlying creativity and creative experience. When our goal is to understand creativity, then instead of solely focusing on the product, we perhaps need to focus on the process (Green et al., 2024; Nath et al., 2024; Holyoak, 2024). In addition, instead of focusing on the outcome, we might want to think about the creative experience (Stein, 1953; Abraham, 2024). Even if the creative output of AI systems and humans might be similar, the process and experience might be vastly different.

**The internal processes underlying creativity**

A central reason for claiming that AI systems do not have similar types of internal processes underlying creativity is the fact that biological systems, such as the brain of a creator, are simply far more complex than any present-day or near-future AI systems. While AI systems, particularly those based on artificial neural networks, may superficially resemble biological brains due to their use of artificial neurons and networks consisting of these neurons, a deeper examination reveals important differences between these systems.

Previous research suggests that creativity involves dynamic interactions of large-scale cortical systems, primarily the default mode network (DMN) and the executive control network (ECN). The DMN, associated with mind-wandering and divergent thinking, is more related to idea generation, while the ECN supports evaluative processes and goal-directed behavior (Beaty et al., 2016). Building on this foundation, a closer examination of the specific architecture and algorithms employed by the brain is necessary to differentiate the internal processing mechanisms underlying artificial and human creativity.

The DMN and ECN are both part of the thalamocortical system which is characterized by a recurrent connectivity not only between cortical areas but also with the nonspecific thalamic nuclei being hubs that coordinate cortical processing (Suzuki et al., 2023). This organization stands in contrast to the architecture of large language models, which typically rely on a feedforward architecture. In these AI models, information flows in a unidirectional manner through layers of transformer decoder blocks (Vaswani et al., 2017), moving from input to output without the intricate feedback loops that are essential in biological brains. In mammalian brains, loops exist at every level of processing, with corticocortical and thalamocortical loops playing a crucial role in cognition (Suzuki et al., 2023; Scott et al., 2024). These loops allow for dynamic and continuous interaction between cortical regions so that activity patterns are not only passed from one processing stage to another but can loop back to previously visited stages both through the thalamus and cortical feedback. The higher-order thalamic nuclei can act as controllers, amplifying and coordinating which cortical areas contribute to ongoing processing (Suzuki et al., 2023). Such recurrent processing might enable the integration of diverse information and iterative refinement of ideas, which some consider central to creativity (Chan & Schunn, 2015; Sawyer, 2021).

Second, in contrast to the transformer architecture, which is relatively homogenous (Vaswani et al., 2017), the complexity of the brain partly lies in the fact that besides the thalamocortical system, it contains several other regions that each have its own learning algorithm (e.g. Caligiore et al., 2019) and thus contribute differently to creative cognition.

For example, the basal ganglia are crucial for skill learning. Creativity is fundamentally dependent on the mastery of specific skills, such as painting, writing, or composing music. These skills are not innate; they are developed over time through practice and are encoded within the brain's cortico-basal ganglia loops (Roth & Ding, 2024). For instance, if the creator is learning how to paint, the skills of holding the brush and moving the hand in the right way involve learning specific sequences of neural activation in these loops (Roth & Ding, 2024). It takes time and practice to internalize these sequences. The basal ganglia are essential for the procedural memory that enables individuals to perform complex tasks effortlessly.

The hippocampus also has a role in theories of creativity (e.g., Benedek et al., 2023) due to its involvement in memory-related processes. The hippocampus also supports prospective simulation through its role in episodic future thinking (Addis et al., 2007), a key aspect of creative problem-solving. Importantly, the hippocampus has quite distinct learning and processing algorithms from the thalamocortical system or the basal ganglia (Kumaran et al., 2016). Hence, the creative brain relies on several different areas that complement each other, while modern AI systems are simpler in their overall architecture. It is possible that future AI architectures will be more inspired by neural architecture, in which case this difference would be smaller, but implementing these specific algorithms in AI is not trivial, as we still do not know precisely how they are implemented in the brain.

Finally, the complexity of biological neurons themselves far exceeds that of their artificial counterparts. Artificial neurons in AI models are simplified units that process inputs and generate outputs based on weighted connections to other neurons. While this model is effective for certain computational tasks, it omits the immense biochemical complexity that characterizes real cells (Ball, 2023). These processes involve intricate molecular interactions, gene expression, and signaling pathways that contribute to the neuron's overall function and its ability to adapt, learn, and respond to the environment (Ball, 2023). Note that this does not mean that intracellular processes directly contribute to creative cognition. Rather, the argument is that the human brain is incredibly complex and it would be naive to think that present-day AI systems somehow miraculously capture the essential internal processes underlying human creativity. In short, as the machinery that supports the internal processes has a different level of complexity, the internal processes of artificial creativity are different from human creativity.

**Creative experience**

In the introduction, it was highlighted that creativity researchers have emphasized the experience of satisfaction as being central to creativity (Rogers, 1954; Stein, 1963; Abraham, 2024). There are also many other types of creative experiences, such as the frustration of being stuck, creative tension, or the joy of finally coming to a solution. As one researcher recently put it: "Human creativity in part depends on consciousness, in the sense of inner experience, notably including felt emotions" (Holyoak, 2024).

Can AI systems have such experiences? The consensus of researchers working on consciousness is that present-day AI algorithms are not conscious (Aru et al., 2023; Butlin et al., 2023; Seth, 2024). Hence, these systems also cannot have creative experiences. These artificial systems do not feel the thrill or the joy and cannot be satisfied by their own work. AI systems are not intrinsically motivated to deliver creative output (Amabile, 1985; Runco, 2023b).

Might artificial intelligence soon become conscious and have such experiences? While there are theories of consciousness that are specified on a relatively abstract level and hence can accommodate artificial consciousness at some point (Butlin et al., 2023), this does not mean that AI systems will soon be conscious. Rather, this might mean that these particular theories of consciousness are not specific enough and do not capture the neurobiological complexity underlying consciousness (Aru et al., 2023). Yet others argue that consciousness is not computational and hence digital AI systems cannot be conscious in principle (Searle, 1992; Penrose, 1994; Koch, 2019; Holyoak, 2024).

In sum, experience matters for human creativity (Stein, 1953; Abraham, 2024; Holyoak, 2024). While it cannot be completely ruled out that AI systems might at some point become conscious, the experience aspect currently separates artificial from human creativity.

**What is the effect of artificial intelligence on human creativity?**

Now, with this background, we can ask, what is the effect of AI on human creativity. One can be optimistic and hope that AI will augment human creativity (Vinchon et al., 2023).

However, there are fundamental reasons to think that AI might distort the internal processes of creativity.

One view is that creativity has always relied on external tools to aid the creative process (Vallée-Tourangeau & Vallée-Tourangeau, 2020), whether it is a notebook for jotting down ideas, a whiteboard for mapping out thoughts, a laptop for drafting work, or now more advanced technologies like generative AI. These tools have traditionally served to support and enhance the internal processes of creativity—helping to organize, refine, and realize creative ideas. At first glance, generative AI might seem like just another tool in this long line of aids, potentially offering even greater benefits by speeding up certain aspects of creative work or providing novel ideas that might not have been considered otherwise (Vinchon et al., 2023). This perspective suggests that generative AI could augment the creative process, leading to enhanced productivity and innovation (Rafner et al., 2023).

However, generative AI differs fundamentally from traditional tools in a critical way: it does not merely support the internal creative processes but has the potential to replace or overtake them. While a notebook or a whiteboard serves as a blank canvas for the creator's ideas, requiring the individual to generate, organize, and refine their thoughts, generative AI can actually perform some of these cognitive tasks itself.

The concern here lies in the potential diminishing of the internal processes of creativity—the very cognitive processes that define and drive human creativity. If generative AI begins to take over these functions, the creative act could shift from being an expression of individual creativity, skill, and knowledge to a process of selecting and refining AI-generated outputs.

Firstly, as explained above, creativity is thought to be fundamentally dependent on the mastery of specific skills. However, as individuals increasingly rely on AI to perform creative tasks, the necessity for developing and internalizing skills diminishes. This is not necessarily an issue when we have a grown-up person who already has the skills necessary to do creative work. However, if children and students grow up with AI constantly helping them out in creative tasks, they might not develop and internalize these skills. This reliance on AI for skill-based tasks may lead to a form of cognitive offloading or outsourcing, where the individual might become less engaged in the process of skill acquisition and, consequently, less capable of performing these tasks independently, i.e., without the AI. Over time, this could result in a reduced ability to engage in creative activities without the assistance of AI, similar to how cognitive offloading in other domains leads to diminished performance in those areas (Risko & Gilbert, 2016).

Secondly, creativity is not just about skill; it also requires the integration of diverse knowledge pieces (Benedek et al., 2023). These knowledge pieces are drawn from an individual's experiences, education, and unique cognitive processes. When a person uses generative AI for creative tasks, there will be a larger contribution of AI-generated knowledge pieces and a reduction in the contribution of these internal knowledge pieces. Hence, the creative output generated by AI, while potentially novel and useful, might be less authentic in that it reflects less of the individual's personal experiences, thoughts, and emotions. This shift raises concerns about the authenticity of creative works produced with significant AI involvement, as they may increasingly reflect the characteristics of the AI rather than the unique identity of the creator. If the outcome is less authentic and authenticity is one of the key differences between human and AI creativity (Runco, 2023a), then the difference between human and AI creativity might diminish – not only because AI systems have improved, but because using these tools decreases human authenticity.

Third, the widespread adoption of the same AI tools by a large number of individuals might pose a significant risk to the diversity and variability of creative outputs. Creativity thrives on the generation of novel and diverse ideas, often fueled by the unique perspectives, experiences, and cognitive processes of individual creators. However, as more people rely on the same AI-driven tools for creative tasks, the final products are likely to become increasingly homogenized and less diverse (Doshi & Hauser, 2024; Laak et al., 2024; Riedl & Bogert, 2024).

Generative AI models are typically trained on vast datasets that encompass a wide range of existing human knowledge and creative works. When many individuals use the same AI tools, they are essentially drawing from the same underlying pool of information and processes. This leads to a convergence of ideas, where the outputs produced by different users become more similar to one another, reducing the overall diversity of creativity. A recent study demonstrated that, indeed, participants generated less semantically distinct ideas when using ChatGPT (Anderson et al., 2024; see also Moon et al., 2024). It has also been shown that the use of generative AI reduces the collective diversity of creative output (Doshi & Hauser, 2024; Moon et al., 2024; Riedl & Bogert, 2024).

As a result, the variability of ideas—the hallmark of human creativity—may shrink. The unique, idiosyncratic internal components that individual creators bring to their work become less pronounced, overshadowed by the more uniform influence of AI. In the long term, this trend could have profound implications for creative fields. If the outputs of AI tools dominate the landscape, there may be less room for truly innovative and disruptive ideas. The richness of human creativity, characterized by a vast array of distinct voices and perspectives, could be diminished as the outputs become more standardized and predictable.

**Conclusion**

The rise of AI has significantly advanced the study of creativity, compelling researchers to rethink traditional definitions. Firstly, the realization that creativity cannot be defined solely by originality and usefulness is crucial. Secondly, these initial efforts to redefine creativity highlight the pressing need to better understand the internal processes—the complex cognitive and neurobiological processes that distinguish human creativity from artificial creativity. Finally, while generative AI offers promising opportunities to enhance human creativity, there are also valid concerns. The potential degradation of essential skills, the decrease in the amount of knowledge pieces that creators contribute based on their own authenticity, and the reduction in idea diversity, suggest that the impact of AI on creativity may not be entirely positive.

**Acknowledgments**

I am grateful to Keith Holyoak, Mykyta Kabrel, Kristjan-Julius Laak, Taavi Luik and Kadi Tulver for their helpful comments. This work was supported by the Estonian Research

Council grant PSG728 and the Estonian Centre of Excellence in Artificial Intelligence (EXAI), funded by the Estonian Ministry of Education and Research.